\documentclass[reprint,aps,physrev,groupedaddress]{revtex4-2}

\usepackage{amsmath}
\usepackage{siunitx}
\usepackage{graphicx}
\usepackage{xcolor}
\usepackage{mathtools}
\usepackage[hidelinks]{hyperref}
\usepackage{lineno}
\usepackage{gensymb}

\newcommand{\pati}[2]{}

\begin{document}


\title{Doppler Pulse Amplification}
\author{Klaas De Kinder}
\author{Amir Bahrami}
\author{Christophe Caloz}
\email{christophe.caloz@kuleuven.be}
\affiliation{Department of Electrical Engineering, KU Leuven, Leuven, Belgium}


\begin{abstract}
    \noindent
    The ability to amplify ultrashort pulses has revolutionized modern laser science, driving advances in various fields such as ultrafast optics and spectroscopy. A pivotal development in this field is chirped pulse amplification (CPA), which stretches, amplifies and recompresses ultrashort optical pulses using dispersive elements to overcome amplification limits. However, CPA faces limitations due to gain narrowing, restricting the final pulse duration. Here, we propose Doppler pulse amplification (DoPA), a novel approach for amplifying ultrashort pulses. While DoPA shares similarities with CPA in that it also stretches, amplifies and recompresses pulses, it differs in how it achieves this temporal compansion. Unlike CPA, DoPA exploits Doppler shifts induced by space-time modulated interfaces through a space-time wedge implementation without chirping. We show that DoPA dynamically shifts the pulse spectrum, effectively mitigating the gain narrowing issue of CPA. Additionally, we show that DoPA enables more compact amplification systems via a space-time Fresnel implementation. This approach may pave the way for more efficient, high-intensity laser systems and expand the potential for applications in both laboratory research and practical environments.
\end{abstract}

\maketitle

      \section{Introduction}
        \pati{Methods to Amplify Ultrashort Optical Pulses \\}{}
        Pulse amplification techniques have continuously evolved to meet the growing demands of various fields, from telecommunications and medical applications to scientific research. Some applications, such as communication~\cite{Bromage2004_Commun_PUB,Keiser2000_Commun_BOOK} and radar systems~\cite{Cook1960_Radar_PUB,Meikle2008_Radar_BOOK}, require only moderate gain amplification. However, others---including laser light-matter interactions~\cite{Mudric2021_Laser_Matter_PUB,Shao_Appl_CPA_PUB}, particle acceleration~\cite{Dangor2002_Accel_Electr_CPA_PUB,Malka2006_Accel_Electr_CPA_PUB}, ultrafast optics~\cite{Shao_Appl_CPA_PUB,Saleh2019_BOOK} and precision eye surgery~\cite{Bron1996_Corneal_CPA_PUB,Hartl_Ultra_Fibre_Lasers_PUB}---demand advanced techniques to amplify ultrashort pulses to very high peak powers. In these cases, directly amplifying ultrashort pulses exceed the saturation level of the amplifier, potentially damaging optical components and introducing nonlinear effects such as self-phase modulation~\cite{Yablonovitch1974_PUB,Stuart1994_SPM_CPA_PUB}. To address this, two main approaches have been developed. The first, chirped pulse amplification (CPA)~\cite{Strickland1985_CPA_PUB,Strickland2019_Nobel_PUB,Mourou1988_PUB,Piskarskas1992_OPCPA_PUB,Li2023_Petawatt_Laser_PUB}, uses dispersive elements, such as diffraction gratings, to temporally stretch the pulse, reducing its peak power and spreading its energy over time. This enables greater amplification before the pulse is recompressed to its original shape and duration, now with significantly higher peak power. Developed by Strickland and Mourou, CPA revolutionized laser science and earned them the 2018 Nobel Prize in Physics~\cite{Strickland1985_CPA_PUB,Strickland2019_Nobel_PUB}. The second approach, divided-pulse amplification (DPA)~\cite{Wise2007_DPA_PUB,Tunnermann2014_Active_DPA_PUB,Richardson2016_DPA_PUB,Tunnermann2017_DPA_PUB,Tunnermann2019_CPA_DPA_PUB,Georges2013_CPA_DPA_PUB,Wise2012_DPA_PUB}, splits the input pulse into multiple identical lower-power copies using a sequence of birefringent crystals. These copies are individually amplified and then recombined, resulting in a pulse with significantly higher peak power.

        \pati{Limitations of Conventional Methods \\}{}
        Despite their success, both approaches come with intrinsic limitations that restrict their amplification performance. On one hand, CPA suffers from gain narrowing, a spectral bottleneck that arises when the spectral width of the pulse exceeds the bandwidth of the gain amplifier, leading to temporal broadening and pulse distortion in the output~\cite{Bado1988_CPA_Limi_PUB,Kapteyn1998_Ultrafast_Laser_PUB,Li2023_Petawatt_Laser_PUB}. Additionally, CPA is constrained by its inability to accommodate the large diffraction grating sizes required for relatively long pulses~\footnote{For a fixed amount of group velocity dispersion and a fixed stretching factor, the size of the diffraction grating scales with the square of the input pulse duration~\cite{Saleh2019_BOOK}---for instance, increasing the pulse duration from $10$~fs to $10$~ps requires enlarging the grating length by a factor of $10^6$---which renders diffraction grating technology impractical~\cite{Richardson2016_DPA_PUB}.}. On the other hand, DPA is limited by the number of pulse divisions, restricting the reduction in peak power of the individual copies. Consequently, amplification is constrained, as the copies quickly reaches the saturation level of the amplifier~\cite{Tunnermann2013_DPA_Lim_PUB,Georges2016_Ultra_Fib_Amp_PUB,Georges2014_DPA_Lim_PUB}.
        
        \pati{DoPA Contribution with USTEMs \\}{}
        We propose a novel method for amplifying ultrashort pulses, called Doppler pulse amplification (DoPA). DoPA shares the same top-level structure with CPA---stretching, amplifying and compressing---but instead of using dispersive elements, it relies on Doppler frequency shifts~\cite{Doppler1842_PUB} induced by moving modulated interfaces~\cite{Gaafar2019_PUB,Caloz2019b_ST_Metamaterials_PUB,Deck-Leger2019_Uni_Vel_PUB,Bahrami2025_Wedges_PUB} to control the pulse compansion (expansion and compression), without any chirping. It mitigates the gain narrowing problem by dynamically alternating the pulse spectrum, making it suitable for the amplification of ultrashort pulses, and may lead to compact systems.

        \pati{Structure Article \\}{}
        The paper is structured as follows. Section~\ref{sec:Doppler_Pulse_Amplification_Concept} introduces DoPA in parallel to CPA. Section~\ref{sec:System_Analysis} analyzes the DoPA system using a space-time wedge implementation. Section~\ref{sec:System_Demonstration} demonstrates this analysis. Section~\ref{sec:Device_Size_Reduction} discusses how the device size can be reduced by replacing the space-time wedge with a space-time Fresnel profile. Section~\ref{sec:Experimental_Outlook} outlines potential experimental realizations of the DoPA system. Finally, Sec.~\ref{sec:Conclusions} concludes the paper. 
        
    \begin{figure*} 
        \includegraphics[width=\linewidth]{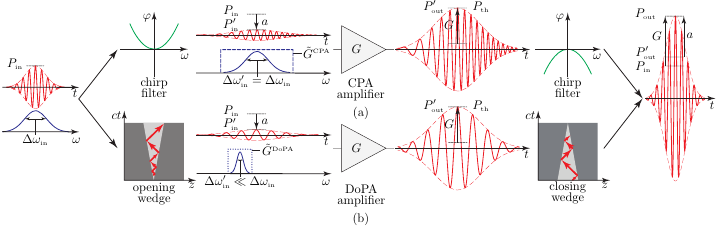}
        \caption{General concept of (a)~chirped pulse amplification (CPA) and (b)~proposed Doppler pulse amplification (DoPA). In both CPA and DoPA, the input pulse is first stretched out in time to reduce its peak power from $P_{\text{in}}$ to $P'_{\text{in}} = P_{\text{in}}/a$, where $a > 1$ is the compansion factor. This temporal stretching---achieved via a chirp filter in CPA and an opening space-time wedge in DoPA---lowers the peak power enough to allow amplification up to the saturation threshold, $P_{\text{th}}$, of the amplifier. With gain, $G$, the amplified stretched pulse reaches $P'_{\text{out}} = GP'_{\text{in}} = P_{\text{th}}$. The pulse is then recompressed---using a matched chirp filter (CPA) and a matched closing space-time wedge (DoPA)---restoring its duration and boosting the peak power to $P_{\text{out}} = aP'_{\text{out}} = GP_{\text{in}}$. Unlike CPA, where the spectral width remains constant through the chirp filter ($\Delta\omega_{\text{in}}' = \Delta\omega_{\text{in}}$)~\cite{Saleh2019_BOOK} and therefore, requires a broad amplifier bandwidth, DoPA dynamically compresses the spectrum during the stretching stage ($\Delta\omega_{\text{in}}' \ll \Delta\omega_{\text{in}}$)~\cite{Bahrami2025_Wedges_PUB}, enabling amplification with a much narrower-band amplifier.}
        \label{fig:CPA_vs_DoPA}
    \end{figure*}

    \section{Concept}\label{sec:Doppler_Pulse_Amplification_Concept}
        \pati{Announcement \\}{}
        Although based on a fundamentally different mechanism, DoPA shares the principle of temporal compansion with CPA. Therefore, it makes sense to introduce DoPA alongside CPA. Figure~\ref{fig:CPA_vs_DoPA} depicts the two approaches, with Fig.~\ref{fig:CPA_vs_DoPA}a and~\ref{fig:CPA_vs_DoPA}b corresponding to CPA and DoPA, respectively.

        \pati{CPA with Chirp Filters \\}{}
        A CPA system, shown in Fig.~\ref{fig:CPA_vs_DoPA}a, consists of three main stages~\cite{Strickland1985_CPA_PUB}. First, the input pulse---with peak power $P_{\text{in}}$---passes through a chirp filter characterized by a phase modulation $\varphi{\left[\omega\right]}$, whose group velocity dispersion temporally expands and chirps the pulse~\cite{Saleh2019_BOOK}. This causes different frequency components to experience different group delays, resulting in temporal stretching of the pulse. This chirping process reduces the peak power from $P_{\text{in}}$ to $P_{\text{in}}/a$, where $a > 1$ is the compansion factor while spreading its energy over a longer duration. Second, the stretched pulse is amplified by a gain amplifier, e.g., Ti:sapphire or Nd:glass, up to its saturation threshold, $P_{\text{th}}$, to mitigate nonlinear effects and prevent damage to the optical components, resulting in an intermediate peak power $P'_{\text{out}} = GP'_{\text{in}}$, where $G$ is the amplifier gain. Finally, a second chirp filter, with opposite dispersion characteristics, compresses the pulse back to its original duration, eliminating the chirp introduced in the first step and significantly increasing the output peak power to $P_{\text{out}} = aP_{\text{out}}' = GP_{\text{in}}$.
        
        \pati{CPA Gain Narrowing Limitation \\}{}
        Since the input pulse is ultrashort, it has a broad spectral width that remains unchanged after passing through the chirp filter~\cite{Saleh2019_BOOK}. This imposes strict constraints on the amplifier, whose limited bandwidth may amplify only a portion of the spectrum, leading to distortions that undermine the original goal of the system. This phenomenon, known as gain narrowing, is one of the principal limitations of CPA. In the figure, it is avoided by employing a very broadband amplifier, but it would arise if the narrower-band amplifier corresponding to the dotted curve in Fig.~\ref{fig:CPA_vs_DoPA}b were used~\cite{Bado1988_CPA_Limi_PUB,Kapteyn1998_Ultrafast_Laser_PUB,Li2023_Petawatt_Laser_PUB}.
        
        \pati{DoPA with Frequency Doppler Shifts \\}{}
        While DoPA also involves sequential pulse expansion, amplification and recompression, as illustrated in Fig.~\ref{fig:CPA_vs_DoPA}b, it achieves these transformations through Doppler frequency shifts induced by moving interfaces, instead of chirp filters. DoPA employs two paired space-time wedges, recently introduced in~\cite{Bahrami2025_Wedges_PUB}. The wedge edges are assumed to act as perfect electric conductor moving interfaces, ensuring that the pulse remains confined within the structure. This configuration allows the pulse to bounce back and forth between two moving interfaces and depending on the wedge type—--either opening or closing—--the temporal envelope of the pulse either expands or compresses upon each interaction with the interfaces. In the case of an opening wedge, the pulse interacts only with comoving interfaces, leading to Doppler downshifting, which stretches the pulse duration by a factor $a$ and reduces its peak power to $P_{\text{in}}/a$, facilitating safe amplification. Conversely, for a closing wedge, the wave encounters only contramoving interfaces, resulting in Doppler upshifting and compressing the pulse back to its original duration and shape. Notably, no chirping occurs here, given the uniformity of the space-time structure~\cite{Bahrami2023_FDTD_PUB,DeKinder2025_Scat_Chirp_ASTEM}.

        \pati{Advantages DoPA Compared to CPA \\}{}        
        This Doppler-driven compansion mechanism causes progressive spectral narrowing as the pulse propagates through the opening wedge (Sec.~\ref{sec:System_Analysis}), in contrast to CPA. This dynamic narrowing enables the use of much narrower-band amplifiers without incurring gain narrowing and preserving spectral integrity after amplification. Additionally, DoPA offers the potential for a compact implementation (Sec.~\ref{sec:Device_Size_Reduction}), which could be particularly advantageous in applications requiring high-power pulses where system size and complexity are critical constraints.

    \section{System Analysis}\label{sec:System_Analysis}
\begin{figure}
            \centering
            \includegraphics[width=\linewidth]{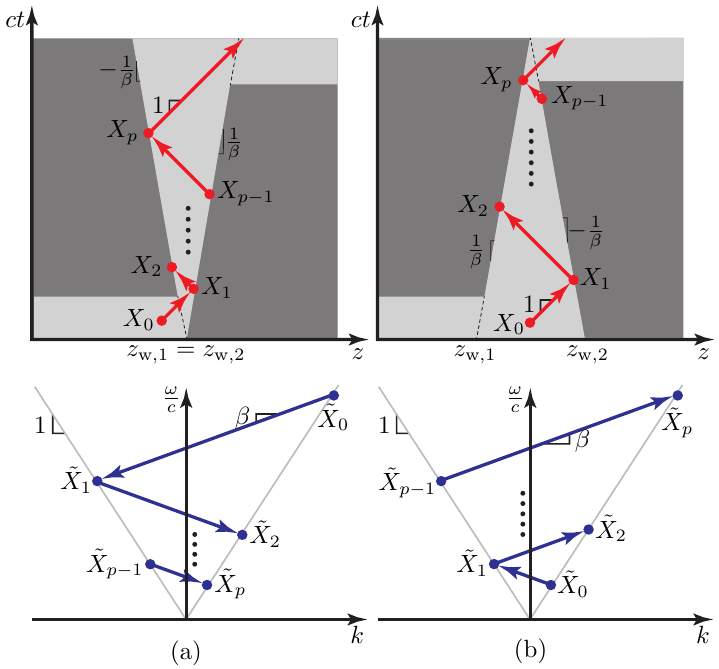}
            \caption{Implementation of the space-time wedges in the DoPA system of Fig.~\ref{fig:CPA_vs_DoPA}b with space-time apertures (top) and corresponding spectral transition diagrams (bottom). 
            The wedges are the light-gray triangular areas, representing propagation media of unit refractive index ($n = 1$), that compand pulses via Doppler shifting [Eq.~\eqref{eq:Doppler_Shift}] by a compansion factor $a$ [Eq.~\eqref{eq:Width_Ratio}]. The two dark-gray areas sandwiching the wedges act as perfect electric conductors. The modulation interfaces travel with normalized velocity $\beta = v_{\text{m}}/c$. The direct space-time points are denoted by $X_{p }= \left(z_{p},ct_{p}\right)$, while the inverse space-time points are denoted by $\tilde{X}_{p} = \left(k_{p},\omega_{p}/c\right)$. (a)~Opening wedge: pulses broaden temporally and compress spectrally. (b)~Closing wedge: pulses compress temporally and broaden spectrally.}
            \label{fig:Example_Wedge}
        \end{figure}

        \pati{Double Wedge Structure \\}{}
        Figure~\ref{fig:Example_Wedge} shows a practical implementation of the double space-time wedge structure of the DoPA system (Fig.~\ref{fig:CPA_vs_DoPA}b) with corresponding spectral transition diagrams. Unlike idealized space-time wedges~\cite{Bahrami2025_Wedges_PUB}, this implementation includes space-time apertures that allow the pulse to enter and exit the structure without spurious reflections. In the opening wedge configuration (Fig.~\ref{fig:Example_Wedge}a), the input aperture is positioned beyond the wedge apex, while the output aperture is before the wedge basis. In the closing wedge configuration (Fig.~\ref{fig:Example_Wedge}b), the input and output apertures correspond to the time-reversed versions of their opening-wedge counterparts. Once inside the structure, the incident pulse---initially at $X_{0} = \left(z_{0},ct_{0}\right)$---undergoes multiple space-time reflections and related spectral transitions at distinct space-time scattering points $X_{1}, X_{2},\hdots$, until its final reflection at $X_{p}$, after which it exits the structure. Both the opening and closing wedges move with a constant normalized modulation velocity, $\beta = v_{\text{m}}/c$, where $0 < \beta < 1$, which governs the Doppler shifts and temporal dynamics of the pulse as it propagates through the structure.    

        \pati{Link with CPA Chirp Filters \\}{}
        To ensure full recovery of the original pulse shape and duration, the opening and closing wedges must be paired with precisely opposite space-time trajectories, i.e., the same velocity magnitude, and an identical number of reflections. This design is analogous to chirp filters in CPA, where the phase modulation in the second filter compensates for the chirp introduced by the first (Fig.~\ref{fig:CPA_vs_DoPA}a). Likewise, in DoPA, the opening wedge modifies the pulse in a controlled manner, while the closing wedge precisely reverses these modifications.

        \pati{Doppler Shift For One Reflection \\}{}
        When the incident pulse impinges on the moving interface, at $X_{1}$, it undergoes the Doppler frequency shift of a moving interface with constant velocity, $\beta$, viz., $\omega_{1} = R\omega_{0}$, where $R = \left(1\mp \beta\right)/\left(1\pm\beta\right)$ and $\omega_{0}$ is the center frequency of the incident wave~\cite{Caloz2019b_ST_Metamaterials_PUB} (Fig.~\ref{fig:Example_Wedge}). The top sign corresponds to an interface that is comoving with respect to the incident wave, resulting in $R < 1$ and thus a decrease in frequency. The bottom sign corresponds to an interface that is contramoving with respect to the incident wave, resulting in $R > 1$ and thus an increase in frequency. Because this scaling applies uniformly to the entire spectrum, both the center frequency, $\omega_{1}$, and the spectral width, $\Delta\omega_{1}$, are multiplied by the same factor, $R$, resulting in spectral compression for comoving interfaces ($R < 1$) and spectral broadening for contramoving interfaces ($R > 1$).

        \pati{Doppler Shift After $p$ Reflections \\}{}
        Within the wedge structure (Fig.~\ref{fig:Example_Wedge}), the pulse experiences multiple space-time reflections, each introducing its own Doppler shift. However, due the uniformity of the structure, the pulse undergoes the same Doppler shift at each scattering event. Therefore, the Doppler-shifted frequency, $\omega_{p}$, after $p$ space-time reflections is simply (Sec.~\ref{Appendix:Doppler_Shift})
        \begin{equation}\label{eq:Doppler_Shift}
            \omega_{p} = R^{p}\omega_{0}\,.
        \end{equation}
        For an opening wedge (Fig.~\ref{fig:Example_Wedge}a), the wave encounters only comoving interfaces ($R < 1)$ and thus both the frequency and spectral width decrease with each reflection. For a closing wedge (Fig.~\ref{fig:Example_Wedge}b), the pulse encounters only contramoving interfaces ($R > 1$) and thus both the frequency and spectral width increase as the wave propagates through the structure.

        \pati{Compansion Factor \\}{}
        In addition to the Doppler shift [Eq.~\eqref{eq:Doppler_Shift}], the temporal envelope of the pulse is affected, exhibiting an inverse relationship with frequency. We define the temporal compansion factor, $a$, as the ratio of the width of the reflected wave, $\tau_{p}$, after $p$ bounces and the temporal width of the incident wave, $\tau_{0}$, viz. (Sec.~\ref{Appendix:Width_Ratio}),
        \begin{equation}\label{eq:Width_Ratio}
            a = \frac{\tau_{p}}{\tau_{0}} = R^{-p}\,.
        \end{equation}
        For an opening wedge, $a > 1$, meaning that the pulse stretches in time, making $a$ a stretching factor. For a closing wedge, $a < 1$, meaning that the pulse compresses, making $a$ a compression factor. Notably, Eq.~\eqref{eq:Doppler_Shift} and Eq.~\eqref{eq:Width_Ratio} exhibit an inverse relationship: compression/expansion in the spectral domain [Eq.~\eqref{eq:Doppler_Shift}] corresponds to expansion/compression in the time domain [Eq.~\eqref{eq:Width_Ratio}] and vice versa.

        \pati{Device Size Wedge Profiles \\}{}
        The physical size of the DoPA system is a key consideration for its practical implementation. The total length of the wedge implementation, $l_{\text{w}}$, depends on the initial pulse duration, $\tau_{0}$, the compansion factor, $a$, and the interface velocity, $\beta$, following the relation (Sec.~\ref{Appendix:Wedge_Profile})
        \begin{equation}\label{eq:Device_Size_Wedge_Profile}
            l_{\text{w}} = a\left(\frac{2}{1+\beta} + \beta\right)c\tau_{0}\,.
        \end{equation}
        For instance, for a femtosecond pulse with $\tau_{0} = \SI{10}{\fs}$, a compansion factor of $a = 10^{6}$ and an interface velocity of $\beta = 0.01$, the required device length is approximately $l_{\text{w}} \approx \SI{6}{\meter}$, which is comparable to the size of a CPA system and requires a complex meandrous structure~\cite{Yakovlev2014_Stretch_Compress_CPA_PUB}. Importantly, there is an inherent trade-off between the interface velocity and device size. At low velocities, the compansion factor per bounce is small, requiring a large number of reflections to achieve the desired compansion. As the interface velocity increases, the compansion factor per bounce grows, reducing the number of required reflections. Notably, this increase in compansion factor per bounce outpaces the reduction in the number of reflections, leading to an overall decrease in device size. This trend continues until a critical velocity, $\beta_{\text{c}} = \sqrt{2} - 1$, at which the device size is minimized, independent of the compansion factor and initial pulse duration. Beyond this point, while the compansion factor per bounce continues to increase---further reducing the number of reflections---the pulse takes longer to catch up with the moving interface, increasing the required size.

        \section{System Demonstration}\label{sec:System_Demonstration}
\begin{figure}
            \centering
            \includegraphics[width=\linewidth]{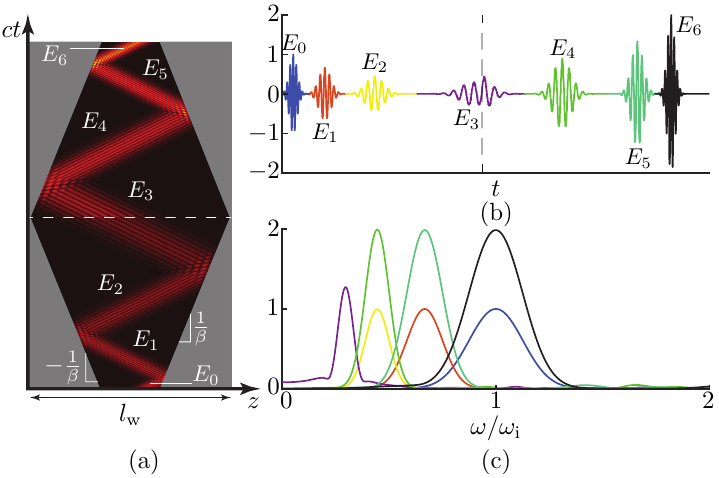}
            \caption{Demonstration of the DoPA system in Fig.~\ref{fig:CPA_vs_DoPA}b with the space-time wedge implementation in Fig.~\ref{fig:Example_Wedge}. (a)~Space-time diagram of the cascaded open-closed wedge structure with a central amplifier. (b)~Time-domain waveforms, with dashed line indicating the amplifying time. (c)~Corresponding spectrum.} 
            \label{fig:Full_DoPA_Wedge}
        \end{figure}

        \pati{Two-Wedge Combination Example \\}{}
        Figure~\ref{fig:Full_DoPA_Wedge} demonstrates the space-time evolution, time-domain waveforms and frequency spectra of a space-time wedge implementation (Fig.~\ref{fig:Example_Wedge}) of the DoPA system (Fig.~\ref{fig:CPA_vs_DoPA}b), governed by the Doppler [Eq.~\eqref{eq:Doppler_Shift}] and temporal width [Eq.~\eqref{eq:Width_Ratio}] scaling relations. Initially, the incident pulse undergoes multiple comoving space-time reflections within the opening wedge, resulting in successive temporal stretching (Fig.~\ref{fig:Full_DoPA_Wedge}b) and Doppler downshifting (Fig.~\ref{fig:Full_DoPA_Wedge}c). Then, the expanded pulse is amplified, at the time indicated by the dashed lines in Fig.~\ref{fig:Full_DoPA_Wedge}a and Fig.~\ref{fig:Full_DoPA_Wedge}b. Finally, the interfaces begin moving toward each other (closing wedge), causing the pulse to undergo multiple contramoving space-time reflections, leading to successive temporal compression and Doppler upshifting. Ultimately, the pulse exits the system with its original duration and shape, but with significantly higher peak power. Notably, the amplifier gain, $G$, in Fig.~\ref{fig:Full_DoPA_Wedge} is only two, due to space constraints, though in practice it can reach $G = 10^{3} - 10^{4}$. 

    \section{Device Size Reduction}\label{sec:Device_Size_Reduction}         
        \pati{From the Wedge to the Fresnel Profile \\}{}
        In the space-time wedge implementation of Fig.~\ref{fig:Full_DoPA_Wedge}, an issue arises due to the continuous motion of the interfaces. As the wedge width increases with the required compansion factor [Eq.~\eqref{eq:Width_Ratio}], its device length [Eq.~\eqref{eq:Device_Size_Wedge_Profile}] may become excessively large. To address this drawback, we adopt the concept of the space-time Fresnel prism~\cite{Li2023_Fresnel_Prism_PUB}. In this approach, the interface positions reset after each reflection, producing an oscillatory, sawtooth-like trajectory.

\begin{figure}
            \centering
            \includegraphics[width=\linewidth]{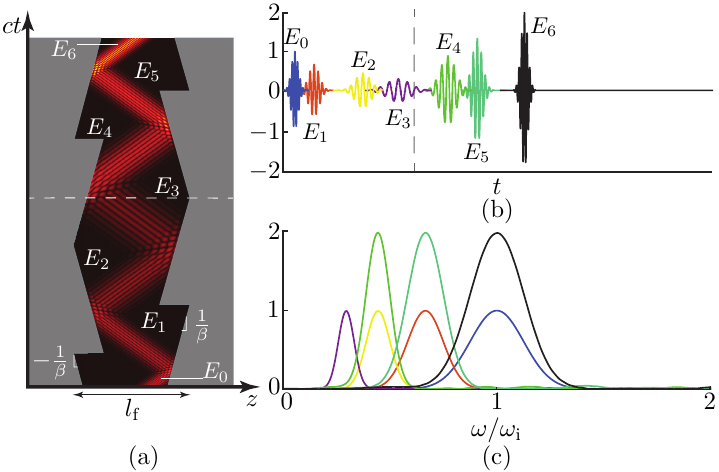}
            \caption{Realization of DoPA (Fig.~\ref{fig:CPA_vs_DoPA}b) using a space-time Fresnel structure with the same number of reflections as in Fig.~\ref{fig:Full_DoPA_Wedge}. (a)~Space-time diagram. (b)~Time-domain waveforms, with the dashed line indicating the amplifying time. (c)~Corresponding spectrum.}
            \label{fig:Full_DoPA_Sawtooth}
        \end{figure}

        \pati{Fresnel Demonstration \\}{}
        Figure~\ref{fig:Full_DoPA_Sawtooth} depicts the space-time evolution, time-domain waveforms and frequency spectra of the space-time Fresnel implementation of the DoPA system (Fig.~\ref{fig:CPA_vs_DoPA}b). The pulse experiences the same sequence of Doppler-induced shifts and compansion as the wedge implementation (Fig.~\ref{fig:Full_DoPA_Wedge}), as evidenced by the similar waveform and spectral evolution. The key difference lies in the oscillatory motion of the interfaces and, hence, the shorter spacings between the temporal pulses, which allows the device length to be significantly reduced. Note that precise synchronization between the two moving boundaries is critical to avoid spurious reflections from the sharp discontinuities. If not carefully managed, these interactions may degrade pulse quality and reduce overall efficiency.

        \pati{Device Size Fresnel Profiles \\}{}
        The device size, $l_{\text{f}}$, of the space-time Fresnel implementation is given by (Sec.~\ref{Appendix:Fresnel_Profile})
        \begin{equation}\label{eq:Device_Size_Fresnel_Profile}
            l_{\text{f}} = 2\left(1 + a\beta\right)c\tau_{0}\,.
        \end{equation}
        For the same values as those below Eq.~\eqref{eq:Device_Size_Wedge_Profile}, Eq.~\eqref{eq:Device_Size_Fresnel_Profile} yields $l_{\text{f}} \approx \SI{6}{\cm}$, which is about a hundred times smaller than the wedge implementation. Unlike Eq.~\eqref{eq:Device_Size_Wedge_Profile}, where the device length scales nonlinearly with the interface velocity, here, it grows linearly. This is because the wavelength of the (periodic) oscillatory motion is directly proportional to the velocity.

\begin{figure}
            \centering
            \includegraphics[width=\linewidth]{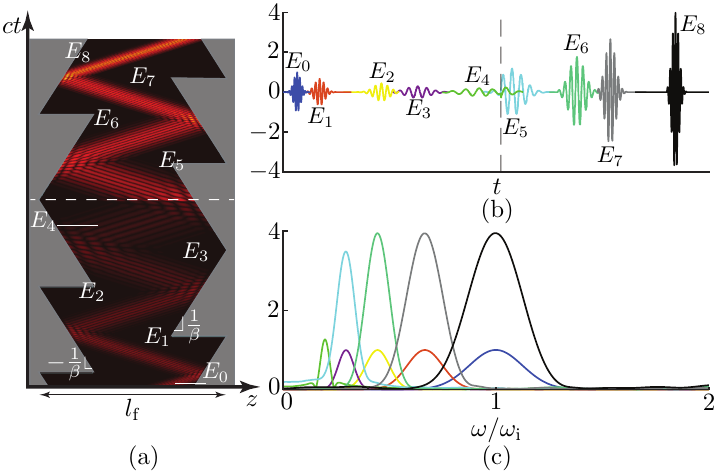}
            \caption{DoPA using a space-time Fresnel structure with the same device size as in Fig.~\ref{fig:Full_DoPA_Wedge} ($l_{\text{f}} = l_{\text{w}}$). (a)~Space-time diagram. (b)~Time-domain waveforms, with the dashed line indicating the amplifying time. (c)~Corresponding spectrum. Note that the dynamic ranges in (b) and (c) are twice those in Figs.~\ref{fig:Full_DoPA_Wedge} and~\ref{fig:Full_DoPA_Sawtooth}.}
            \label{fig:Full_DoPA_Sawtooth_Same_Length}
        \end{figure}

        \pati{Example Same Device Size \\}{}
        The contrast in device size between the wedge and Fresnel configurations is evident when comparing Figs.~\ref{fig:Full_DoPA_Wedge} and~\ref{fig:Full_DoPA_Sawtooth}. To further illustrate this, Fig.~\ref{fig:Full_DoPA_Sawtooth_Same_Length} shows the case when $l_{\text{f}} = l_{\text{w}}$. In this scenario, the Fresnel profile allows for an additional expansion-compression pair, further reducing the peak power of the pulse. This enables a larger amplification gain---four in this case---before reaching the saturation threshold of the amplifier.

    \section{Experimental Outlook}\label{sec:Experimental_Outlook}
        \pati{Implementation}{
        }

        The DoPA system hinges on the ability to generate interfaces moving at relatively high speeds, posing significant experimental challenges. Although no direct method currently exists for creating such interfaces, recent advances in time-varying media~\cite{Alu2023_Temp_Refl_PUB,Alu2023_Coherent_Wave_Control_PUB,Peroulis2024_Time_Ref_TEM_PUB,Boyd2020_Time_Refr_ENZ_TEM_PUB,Segev2023_Single_Cycle_TEM_PUB,Kinsey2025_ST_Knife_TEM_PUB} provide a foundation that could be extended to space-time varying systems. Here, we outline how these existing platforms might be adapted to demonstrate a proof-of-concept realization of DoPA.

        \pati{Microwave Regime}{
        }

        In the microwave regime, pure time interfaces have been realized by temporally modulating the effective impedance of artificial transmission lines, controlled via high-speed electronic components, such as electronic switches~\cite{Alu2023_Temp_Refl_PUB,Alu2023_Coherent_Wave_Control_PUB} or high-speed photodiodes~\cite{Peroulis2024_Time_Ref_TEM_PUB}. Extending this concept to space-time systems requires sequential, spatially coordinated switching to the ground that emulates a moving perfect electric conducting (PEC) interface, demanding precise spatial control of the switching signals. One strategy is to drive the switches using a field-programmable gate array (FPGA), allowing for precise tunable delays and sequencing. By synchronizing the activation of switches in a controlled  pattern, it is possible to form two contramoving interfaces along the line, effectively realizing space-time wedges. Microwave implementations employing corporate feed networks with tailored optical delays offer another route.  The key DoPA parameter is the normalized modulation velocity, $\beta = \Delta x / \left(c\Delta t\right)$, where $\Delta x$ is the distance between adjacent switching elements and $\Delta t$ is the time delay between their activation. To emulate sharp moving interfaces corresponding to the wedges, two conditions must be met. First, the local rise time at each site must remain shorter than $\Delta t$, ensuring that the wave perceives a distinct discontinuity. Second, $\Delta x$ must be small compared to the wavelength of the incoming pulse to ensure a smooth modulation profile. For instance, using parameters from~\cite{Alu2023_Coherent_Wave_Control_PUB}: $\Delta x = 0.208$~m and $\Delta t = 4$~ns yield $\beta = 0.17$. With $\tau_{0} = 0.5$~ns and $a = 10^{2}$, the Fresnel device length [Eq.~\eqref{eq:Device_Size_Fresnel_Profile}] equals $l_{\text{f}} \approx 5.5$~m, which is smaller than the $6.24$~m transmission line used in~\cite{Alu2023_Temp_Refl_PUB,Alu2023_Coherent_Wave_Control_PUB}.

        \pati{Optical Regime}{
        }

        Experiments in the optical regime have also demonstrated temporal discontinuities using pump-probe setups~\cite{Boyd2020_Time_Refr_ENZ_TEM_PUB,Segev2023_Single_Cycle_TEM_PUB,Kinsey2025_ST_Knife_TEM_PUB}, which induce refractive index changes via intense pump pulses in nonlinear media. The main challenge lies in achieving index variations that are both sufficiently large and fast. Transparent conducting oxides, especially in the epsilon-near-zero regime, have emerged as promising candidates, capable of refractive index modulations approaching unity on femtosecond timescales. Space-time modulated configurations can be achieved by tilting the pump beam with respect to the probe beam~\cite{Caloz2019b_ST_Metamaterials_PUB}. Due to the oblique geometry, the pump-induced refractive index change sweeps across the medium, effectively forming a moving interface. Using two contratilted pump beams allows the creation of a space–time wedge structure. For instance, if the angle between the probe and pump pulse is $60\degree$ and the pump is traveling at the speed of light, the normalized modulation velocity is given by $\beta = 1/\tan 60\degree = 1/\sqrt{3}$. Next, taking $a = 10^{3}$ and $\tau_{0} = 10$~fs yields a Fresnel length~$l_{\text{f}} \approx 3.5$~mm.

    \section{Conclusions}\label{sec:Conclusions}
        \pati{Summary \\}{}
        We introduced Doppler pulse amplification (DoPA), a novel approach that harnesses space-time modulation interfaces to amplify ultrashort pulses via Doppler frequency shifts, complementing established techniques like chirped pulse amplification (CPA) and divided-pulse amplification (DPA). We analyzed the system with a space-time wedge implementation, demonstrating how multiple space-time scattering events and associated Doppler frequency shifts can be utilized to achieve compansion, while also mitigating the gain narrowing problem inherent in CPA. Furthermore, we explored a space-time Fresnel architecture approach, which reduces system size without sacrificing amplification efficiency, making DoPA suitable for amplification of ultrashort pulses. This compact design makes DoPA particularly suitable for applications requiring high-power ultrashort pulses, especially where system size and complexity are critical factors. Finally, we outlined potential experimental realizations of the DoPA system in both the microwave and optical regimes.

    \noindent\textbf{Author contributions} \\
    K. De Kinder performed the bulk of the work. A. Bahrami assisted with the calculations. C. Caloz has supervised the work. All authors have accepted responsibility for the entire content of this manuscript and approved its submission.    
    
    \noindent\textbf{Competing interests} \\
    The authors declare no competing interests.

\appendix
    \section{Space-Time Wedge Scattering Solution}\label{Appendix:Space-Time_Wedge_Scattering_Solution}
        \subsection{Scattering Solution}
            \pati{Scattering Solution \\}{}
            The scattering solution, $E_{\text{w}}$, for an electromagnetic wave inside a space-time perfect electric conductor wedge with both interfaces moving at a constant normalized velocity $\beta$ is a particular case of the expressions in Eq.~(5) of~\cite{Bahrami2025_Wedges_PUB}, with $v_{1} = \mp\beta c$, $v_{2} = \pm\beta c$ and $u = c$, where the top sign refers to an opening space-time wedge and the bottom sign to a closing wedge. The initial positions of the interfaces are denoted by $z_{\text{w},1}$ and $z_{\text{w},2}$. The solution is a superposition of forward and backward waves and is given by
            \begin{equation}\label{eq:appendix:General_Solution_Electromagnetic_Wave_Wedge}
                \begin{split}
                    E_{\text{w}} &=\sum_{p=2k}R^{p}E_{\text{i}}\underbrace{{\left[\Delta\phi_{k-1}^{-} + R^{p}\left(\frac{z}{c} -t\right)\right]}}_{\Phi_{p}} \\
                    &-\sum_{p=2k+1}R^{p}E_{\text{i}}{\left[\Delta\phi_{k}^{+} - R^{p}\left(\frac{z}{c}+t\right)\right]}\,,
                \end{split}
            \end{equation}
            where $E_{\text{i}}$ is the incident wave,
            \begin{equation}\label{eq:appendix:Definition_R}
                R = \frac{1\mp\beta}{1\pm\beta}\,,
            \end{equation}
            and $p$ denotes the number of scattering events. If $p$ is even, the solution represents a forward wave while if $p$ is odd, it represents a backward wave. The phase shifts in Eq.~\eqref{eq:appendix:General_Solution_Electromagnetic_Wave_Wedge} are defined as 
            \begin{equation}
                \begin{split}
                    \Delta\phi_{k}^{+} &= A - R\frac{1-R^{2k}}{1-R^{2}}B\,, \\
                    \Delta\phi_{k}^{-} &= \Delta\phi_{k}^{+} - R^{2k + 1}C\,, 
                \end{split}
            \end{equation}
            where
            \begin{equation}
                \begin{split}
                    A &= \left(1+R\right)\frac{z_{\text{w},2}}{c} - R\phi_{0}^{-} - \phi_{0}^{+} \,, \\
                    B &= \left(1+R\right)\frac{z_{\text{w},1} - Rz_{\text{w},2}}{c} - \left(1 - R^{2}\right)\phi_{0}^{-} \,, \\
                    C &= \left(1 + R\right)\frac{z_{\text{w},1}}{c} - R\phi_{0}^{+} - \phi_{0}^{-}\,.
                \end{split}
            \end{equation}
            The initial phase shifts are defined as
            \begin{equation}
                \phi_{0}^{\pm} = \frac{z_{0}}{c} \mp t_{0} \,.
            \end{equation}

    \subsection{Doppler Shift}\label{Appendix:Doppler_Shift}
        \pati{Doppler Shift \\}{}
        The shifted frequency, $\omega_{p}$, after $p$ bounces may be calculated by taking a time derivative of the phase argument in Eq.~\eqref{eq:appendix:General_Solution_Electromagnetic_Wave_Wedge}:
        \begin{equation}\label{eq:appendix:Doppler_Shift}
            \omega_{p} = \frac{\partial\Phi_{p}}{\partial t}\omega_{0} = R^{p}\omega_{0} \,.
        \end{equation}

    \subsection{Compansion Factor}\label{Appendix:Width_Ratio}
        \pati{Compansion Factor \\}{}
        In addition to the frequency Doppler shift [Eq.~\eqref{eq:appendix:Doppler_Shift}], the temporal envelope of the pulse also undergoes modification. Due to the Fourier limit, the width of the optical pulse changes by a factor $R^{-p}$. Thus, the compansion factor, $a = \tau_{p} / \tau_{0}$, defined as the ratio of the width of the reflected wave, $\tau_{p}$, and the width of the incident wave, $\tau_{0}$, is given by
        \begin{equation}\label{eq:appendix:Width_Ratio}
            a = \frac{\tau_{p}}{\tau_{0}} = R^{-p}\,.
        \end{equation}

    \section{Device Size for a Given Number of Space-Time Reflections}\label{Appendix:Device_Size}
        \subsection{Wedge Profile}\label{Appendix:Wedge_Profile}
            \pati{Preliminaries \\}{}
            In this section, we determine the minimal size of the wedge profile (Fig.~\ref{fig:Example_Wedge}) required for the compansion factor [Eq.~\eqref{eq:Width_Ratio}] to equal $a$ after $p$ bounces. We perform the analysis for the pulse expansion (first step in DoPA, Fig.~\ref{fig:CPA_vs_DoPA}), i.e., an opening space-time wedge (Fig.~\ref{fig:Example_Wedge}a). The two edges of the space-time wedge may be parametrized as
            \begin{equation}\label{eq:appendix:Parametrization_Legs_Interface}
                \begin{split}
                    z_{\text{L}}{\left[t\right]} &=  -\beta ct - z_{\text{w},0}\,, \\
                    z_{\text{R}}{\left[t\right]} &= \phantom{+}\beta ct + z_{\text{w},0}\,, 
                \end{split}
            \end{equation}
            where we assume a symmetric setup around the $ct$-axis and $z_{\text{L}}{\left[t\right]}$ represents the left edge and $z_{\text{R}}{\left[t\right]}$ the right edge of the wedge. To ensure that the initial pulse can pass through the space-time aperture of the wedge, we require that the initial width of the wedge, given by $z_{\text{R}}{\left[0\right]} - z_{\text{L}}{\left[0\right]} = 2z_{\text{w},0}$, must be at least as large as twice the initial width of the incident pulse. This leads to the condition
            \begin{equation}\label{eq:appendix:Initial_Width_Condition_Wedge_Profile}
                z_{\text{w},0} \geq c\tau_{0}\,,
            \end{equation}
            where we will assume equality for minimal device size. The trajectory of the reflected electromagnetic pulse after $i$ bounces is parametrized as:
            \begin{equation}\label{eq:appendix:Parametrization_Reflected_Wave_i_Bounces}
                z_{i}{\left[t\right]} = \left(-1\right)^{i}c\left(t-t_{i}\right) + z_{i}\,,
            \end{equation}
            where $t_{i}$ and $z_{i}$ are the scattering time and position respectively (Fig.~\ref{fig:Example_Wedge}), with $t_{0} = 0 = z_{0}$. When $i$ is even, Eq.~\eqref{eq:appendix:Parametrization_Reflected_Wave_i_Bounces} represents a forward wave and when $i$ is odd, a backward wave [Eq.~\eqref{eq:appendix:General_Solution_Electromagnetic_Wave_Wedge}].

            \pati{Derivation Device Size Wedge Profile \\}{}
            We are seeking the location of the $p$-th bounce, i.e., $z_{p} = z_{p-1}{\left[t_{p}\right]}$ and shall proceed iteratively. The position, $z_{1}$, of the first reflection is the intersection of the world line of the incident wave [Eq.~\eqref{eq:appendix:Parametrization_Reflected_Wave_i_Bounces}] and the world line of the right edge of the wedge [Eq.~\eqref{eq:appendix:Parametrization_Legs_Interface}], i.e., $z_{0}{\left[t_{1}\right]} = z_{\text{R}}{\left[t_{1}\right]}$, which can be solved for $t_{1}$ as
            \begin{equation}\label{eq:appendix:First_Scattering_Time_Wedge}
                t_{1} = \frac{1}{1-\beta}\tau_{0}\,.
            \end{equation} 
            The position of the first space-time reflection is then given by 
            \begin{equation}\label{eq:appendix:First_Scattering_Position_Wedge}
                z_{1} = z_{\text{R}}{\left[t_{1}\right]} = \frac{1}{1-\beta}c\tau_{0}\,.
            \end{equation}
            The position of the second reflection, $z_{2}$, is the intersection of the world line of the first reflected wave [Eq.~\eqref{eq:appendix:Parametrization_Reflected_Wave_i_Bounces}], with the first scattering time $t_{1}$ [Eq.~\eqref{eq:appendix:First_Scattering_Time_Wedge}] and position $z_{1}$ [Eq.~\eqref{eq:appendix:First_Scattering_Position_Wedge}], and the left leg of the wedge interface [Eq.~\eqref{eq:appendix:Parametrization_Legs_Interface}], i.e., $z_{1}{\left[t_{2}\right]} = z_{\text{L}}{\left[t_{2}\right]}$, which can be solved for $t_{2}$, yielding
            \begin{equation}
                t_{2} = \frac{3-\beta}{\left(1-\beta\right)^{2}}\tau_{0}\,.
            \end{equation}
            Hence, the second scattering event position is given by
            \begin{equation}
                z_{2} = z_{\text{L}}{\left[t_{2}\right]} = -\frac{1+\beta}{\left(1-\beta\right)^{2}}c\tau_{0}\,.
            \end{equation}
            This process repeats. By induction, we find that the time and position of the $p$-th bounce are given by
            \begin{equation}\label{eq:appendix:Expression_Position_and_Location_p_Bounce}
                \begin{split}
                    t_{p} &= \frac{1}{\beta}\left(\frac{a}{1+\beta} - 1\right)\tau_{0} \,, \\
                    z_{p} &=  \left(-1\right)^{p}\frac{a}{1+\beta}c\tau_{0} \,,
                \end{split}     
            \end{equation}
            where the compansion factor is given by Eq.~\eqref{eq:Width_Ratio}. Equation~\eqref{eq:appendix:Expression_Position_and_Location_p_Bounce} refers to the scattering time and position of the \emph{center} of the pulse. At this point, the wave has already expanded to a temporal width $a\tau_{0}$ and thus the final part of the expanded wave extends $a\tau_{0}/2$ above the center. Consequently, the last part of the wave reaches the wedge at time $t_{p} + a\tau_{0}/2$ and needs to be taken into account. Therefore, we compute $z_{\text{L,R}}{\left[t_{p} + a\tau_{0}/2\right]}$ using Eq.~\eqref{eq:appendix:Parametrization_Legs_Interface}, viz.,
            \begin{equation}\label{eq:appendix:Position_Stretched_Out_Wave}
                z_{\text{L,R}}{\left[t_{p} + \frac{a}{2}\tau_{0}\right]} = \frac{\left(-1\right)^{p}}{2}\left(\frac{2}{1+\beta} + \beta\right)ac\tau_{0} \,.
            \end{equation}
            The device size, $l_{\text{w}}$, is twice the value in Eq.~\eqref{eq:appendix:Position_Stretched_Out_Wave}, i.e.,
            \begin{equation}\label{eq:appendix:Device_Length_Wedge}
                l_{\text{w}} = a\left(\frac{2}{1+\beta} + \beta\right)c\tau_{0}\,.
            \end{equation}

        \subsection{Fresnel Profile}\label{Appendix:Fresnel_Profile}
            \pati{Derivation Device Size Fresnel Profile \\}{}
            The space-time Fresnel trajectory can be parametrized as a sawtooth profile with period $T$ and constant velocity $\beta$. The characteristic wavelength of this profile, $\lambda_{\text{f}}$, is given by
            \begin{equation}\label{eq:appendix:Wavelength_Fresnel_Profile}
                \lambda_{\text{f}} = \beta c T\,.
            \end{equation}
            The total device length, $l_{\text{f}}$, consists of two contributions: (i)~twice the distance traveled by the interface in a single oscillation [Eq.~\eqref{eq:appendix:Wavelength_Fresnel_Profile}] and (ii)~the fixed separation between the two edges. Here, we assume this separation to be the same as in the space-time wedge profile [Eq.~\eqref{eq:appendix:Initial_Width_Condition_Wedge_Profile}]. This leads to
            \begin{equation}\label{eq:appendix:Device_Length_First_Observation_Sawtooth}
                l_{\text{f}} = 2\beta cT + 2c\tau_{0} \,.
            \end{equation}
            To ensure that the expanded wave remains confined within a single oscillation and does not interact with the sharp edges, we impose the condition
            \begin{equation}
                T \geq a\tau_{0} \,.
            \end{equation}
            Substituting this constraint into Eq.~\eqref{eq:appendix:Device_Length_First_Observation_Sawtooth}, we obtain the final expression for the device length of the Fresnel trajectory profile:
            \begin{equation}
                l_{\text{f}} = 2\left(1 + a\beta\right)c\tau_{0} \,.
            \end{equation}

\bibliography{DoPA}

\end{document}